# The sufficiently smart compiler is a theorem prover

Extended abstract

Joachim Breitner
University of Pennsylvania
3330 Walnut Street
Philadelphia, PA 19104, USA
joachim@cis.upenn.edu

## ABSTRACT

That the Haskell Compiler GHC is capable of proving non-trivial equalities between Haskell code, by virtue of its aggressive optimizer, in particular the term rewriting engine in the *simplifier*. We demonstrate this with a surprising little code in a GHC plugin, explains the knobs we had to turn, discuss the limits of the approach and related applications of the same idea, namely testing that promises from Haskell libraries with domain-specific optimizations hold.

## 1 INTRODUCTION

Functional programming languages are advertised as well suited to formal reasoning. Nevertheless, a programmer who actually tries to rigorously reason that one function definition in, say, Haskell is equivalent to another receives no support his development tools and has to resort to manual, handwritten proofs or external tools.

This is unfortunate, as the one tool that every programmer uses, and which really knows what a program means, can actually do some of these proofs: The compiler!

If we assume the compiler to be sufficiently smart, then it will optimize any piece of source code to the "best" output code possible. It follows immediately that two semantically equivalent expressions will be compiled to the same output. So to prove that two expressions are equivalent, we just have to ask the compiler to compile both, compare the output, and if they are the same we know that the input expressions are the same. Done.

If this paragraph made you raise your eyebrows, you can lower them again: We are well aware that this bold approach poses a few questions:
  (1) How do we "ask the compiler" in the first place? Can we include this in the user's program workflow?
  (2) How do we compare the output?

  (3) The sufficiently smart compiler does not exist. Will the actually existing compiler be able to prove any interesting equivalences?
  (4) If the compiler says that it cannot prove two expressions to be equal, what, if anything, does that mean?
  (5) If the compiler says that it can prove the two expressions to be equal, what does *that* mean, given that compilers have bugs too?
  (6) Besides proofs of program equivalence, is there anything else we can do with this idea?

Our contributions are raising these questions, and giving preliminary answers, which we summarize here:
  (1) Custom syntax, embedded in the actual code, would be most convenient, but even without extending the language (and hence the whole compiler pipeline, starting with the parser) it is possible to include the proof obligations in the Haskell code, to be picked later in the pipeline by a compiler plugin. This way, no new or additional tools have to be employed.
  (2) There is no need to compare machine code: Almost all interesting optimizations are applied by GHC's simplifier, which implements a rewrite engine. It suffices to compare simplified terms in GHC's intermediate language Core.
  (3) Yes. We proved the Functor, Applicative and Monad laws of a simple, but already non-trivial data structure, as well as a number of rewrite rules from hlint's database.
  (4) It certainly does not mean that the expressions are not equivalent. But inspecting the differences in the simplified expressions can still provide useful insights, such as differences in strictness.
  (5) It does mean that the expressions are equivalent. In a not completely tongue-in-cheek sense the only semantics for Haskell that matters in practice is the one defined by how the compiler compiles a program. Its proofs are therefore vacuously correct.
  (6) The author of a Haskell library who carefully crafts it so that certain code (e.g. composition of list-processing functions) is optimized to more efficient, less pleasant code (e.g. explicit recursion) can use the same technique in his normal library test suite, ensuring that his promises continue to hold, even as the compiler is upgraded to newer versions.

### 1.1 Pedestrian Proofs

Consider a Haskell library author who wants to provide instances for the Functor, Applicative and Monad type classes. He would not

```
data Succs a = Succs a [a]

getCurrent :: Succs t → t
getCurrent (Succs x _) = x

getSuccs :: Succs t → [t]
getSuccs (Succs _ xs) = xs

instance Functor Succs where
    fmap f (Succs o s) = Succs (f o) (map f s)

instance Applicative Succs where
    pure x = Succs x [ ]
    Succs f fs ⟨∗⟩ Succs x xs
        = Succs (f x) (map ($x) fs ++ map f xs)

instance Monad Succs where
    Succs x xs ⋙ f = Succs y (map (getCurrent ∘ f) xs ++ ys)
        where Succs y ys = f x
```

**Figure 1: A non-trivial functor with type class instances**

```
  pure (∘) ⟨∗⟩ Succs u us ⟨∗⟩ Succs v vs ⟨∗⟩ Succs w ws
= Succs (∘) [ ] ⟨∗⟩ Succs u us ⟨∗⟩ Succs v vs ⟨∗⟩ Succs w ws
= Succs (u∘) (map (∘) us) ⟨∗⟩ Succs v vs ⟨∗⟩ Succs w ws
= Succs (u ∘ v) (map ($v) (map (∘) us) ++ map (u∘) vs)
    ⟨∗⟩ Succs w ws
= Succs (u ∘ v) (map (($v) ∘ (∘)) us ++ map (u∘) vs)
    ⟨∗⟩ Succs w ws
= Succs ((u ∘ v) w) (map ($w) (map (($v) ∘ (∘)) us
    ++ map (u∘) vs) ++ map (u ∘ v) ws)
= Succs ((u ∘ v) w) (map (($w) ∘ ($v) ∘ (∘)) us
    ++ map (($w) ∘ (u∘)) vs ++ map (u ∘ v) ws)
= Succs (u (v w)) (map (λu. u (v w)) us
    ++ map (λv. u (v w)) vs ++ map (λw. u (v w)) ws)
= Succs (u (v w)) (map ($(v w)) us
    ++ map u (map ($w) vs ++ map v ws))
= Succs u us ⟨∗⟩ Succs (v w) (map ($w) vs ++ map v ws)
= Succs u us ⟨∗⟩ (Succs v vs ⟨∗⟩ Succs w ws)
```

**Figure 2: A manual proof of the second Applicative law**

only be expected to implement these methods, as shown in Figure 1, but also ensure that these instances fulfill the laws of the type class. He can do so by testing (and indeed, property-based testing [2] would nicely work here) or proving the desired properties; in this paper we focus on the latter.

If the user had written this code in the language of a theorem prover such as Gallina, then he could simply prove if in that system. Alas, Haskell is not a theorem prover, so what are the options? Certainly he could re-implement the code in a different system and perform the proofs there (and maybe the translation can be automated to some extent [6, 7]), but that requires knowing the other system first, and involves quite some extra work. So especially with smaller proof tasks such as this one, the author is likely to do the proofs "by hand", maybe in a comment of the file, and write long chains of equations as in Figure 2.

### 1.2 The Compiler is in the Know

Assume the programmer had added functions to the module that represent the left-hand-side and the right-hand-side of the one of the proof obligations, as follows:

```
functor_law1_rhs :: Succs a → Succs a
functor_law1_rhs x = x
functor_law1_lhs :: Succs a → Succs a
functor_law1_lhs x = fmap id x
```

What will the compiler make of that? We can view the intermediate code using `-ddump-simpl` and list the relevant lines:

```
$ ghc -O -ddump-simpl Succs.hs|grep functor_law1_lhs
functor_law1_lhs :: forall a. Succs a -> Succs a
functor_law1_lhs = functor_law1_rhs
```

This means that the compiler already knows that the left-hand side and the right-hand side of this equations are the same! How is it that it knows that, and why does this equation appear in the code?

We invoked GHC with the `-O` flag, telling it to optimize the code. Three optimizations play a role in this example:

- *Inlining* [4]: The call to fmap is replaced with the definition in Figure 1, and we obtain
  functor_law1_lhs =
      λx. **case** x **of** Succs o s → Succs (id o) (map id s).
  Further inlining replaces id o with o.
- *Rewrite rules* [5]: The base library code contains a number of rewrite rules which allow the compiler to know the identity map id xs = xs. Therefore, we obtain
  functor_law1_lhs =
      λx. **case** x **of** Succs o s → Succs o s.
- *Common subexpression elimination*: Given this code, the compiler knows that the case analysis is redundant, and simplifies this function definition further to
  functor_law1_lhs = λx. x.
  
  But this is precisely the definition of functor_law1_rhs! So, again by common subexpression elimination, the compiler lets both names refer to the same function:
  functor_law1_lhs = functor_law1_rhs

At this point, the programmer feels silly. Why did he spend the effort of manually proving these equations when they are already known by the tool that he uses already, namely the compiler?

## 2 PROOF BY COMPILATION

To get rid of this silliness we need to provide a way for the user to declare equations that he wants the compiler to check, and then actually check them.

A slightly simplified view of the compiler pipeline is shown in the top line of Figure 3; the most relevant component for us is the simplifier in the middle: It receives the program in GHC's intermediate language Core to simplify, optimize and rewrite it. Its output (dumped via `-ddump-simpl`) is where we discovered the statement of the desired equality.

So one approach would be to create a dedicated tool based on the compiler code (or ghc-the-library) runs this pipeline only until the simplifier and then checks if the desired equations hold. But one goal here is to *not* require the programmer to reach out to a new tool,

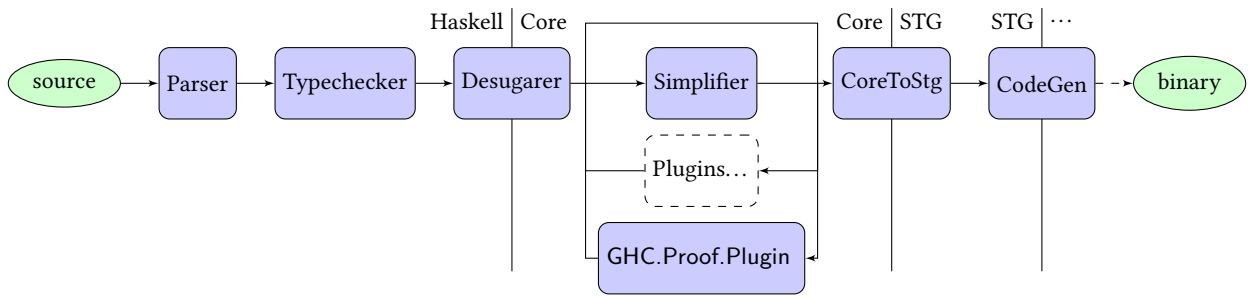

Figure 3: The GHC pipeline and the plugin hook

```
module GHC.Proof where
data Proof = Proof
(===) :: a → a → Proof
_ === _ = Proof
infix 0 ===
{-# INLINE [0] (===) #-}
```

Figure 4: The proof obligation combinator

and we have to use the existing hooks provided by the compiler. One such hook is provided in the form of *GHC plugins*, which are dynamically loaded additional core-to-core passes functions performed by GHC.

## 2.1 Syntax for proof obligations

Unfortunately, the plugin interface is intended to introduce additional optimization passes, not new language features, and does not allow use to define new syntax for the proof obligations. But we can work around that limitation: The module GHC.Proof shown in Figure 4 exports the operator (===) whose type allows it to relate any two expressions of the same type.

A proof obligation is now simply an occurrence of (===) bound to a top-level name. The return type of the operator allows us to give descriptive type signatures to these names:

```
{-# OPTIONS_GHC -O -fplugin GHC.Proof.Plugin #-}
import GHC.Proof
...
functor_law1 :: Succs a → Proof
functor_law1 = fmap id x === x
```

The actual implementation is irrelevant and is chosen so that at the end of the compilation pipeline (===) has been inlined and its arguments have been removed as dead code, so that embedded proof obligations do not increase the size of the compiled program.

## 2.2 The compiler plugin

The GHC flag `-fplugin GHC.Proof.Plugin`, which we can embed in the file as shown above, instructs the compiler to load our plugin, which inserts itself early in the optimization pipeline, directly after the desugarer. It

(1) scans the code for occurrences of lhs === rhs,

(2) passes lhs to the simplifier to obtain its optimized counterpart lhs', and likewise for rhs,
(3) compares lhs' and rhs' for alpha-equality.
(4) If all such pairs are alpha-equivalent, the plugin prints an encouraging message and compilation continues:
```
$ ghc -O Succs.hs
[1 of 1] Compiling Succs ( Succs.hs, Succs.o )
GHC.Proof: Proving functor_law1 ...
GHC.Proof proved 1 equalities
```
(5) But if they are not alpha-equivalent, e.g. trying to prove x ⟨$ y === y, the plugin prints the differing left- and right-hand sides of the equation and aborts compilation:
```
$ ghc -O Succs.hs
[1 of 1] Compiling Succs ( Succs.hs, Succs.o )
GHC.Proof: Proving functor_law1 ...
GHC.Proof: Proving wrong_law ...
Proof failed
  Simplified LHS:
    case x of { Succs o s ->
    Succs.Succs @ a y
        (map @ a @ a ( _ [Occ=Dead] -> y) s)
    }
  Simplified RHS: x

Succs.hs: error: GHC.Proof could not
  prove all equalities
```
Since the plugin works at the stage of Core, these expressions are necessarily Core expressions and include, for example, explicit type applications (Succs @ a).

All heavy lifting is done by existing parts of the compiler (simplification and checking for alpha-equality), so the plugin itself simple and short (< 100 lines of code).

## 2.3 Gory implementation details

Nevertheless, there are a few technicalities that the plugin has to get right for this to work:

*2.3.1 Pipeline location.* A GHC compiler plugin inserts itself into the long list of Core-to-Core transformations performed by GHC. Ideally, we would like to install our plugin first, so that we can get our hands on the proof in their most pristine form and have full control over the optimizations. Unfortunately, the output of the desugarer (cf. Figure 3) is still incomplete. For example, important information about a function defined in the current module is not yet available at the use-site of that function, including information required to inline that function. But without inlining, many proofs do not go through.

The first Core-to-Core pass is a "gentle" run of the simplifier, which takes care of this. Therefore, we insert the plugin as the second pass, right after that.

*2.3.2 Optimizations to run.* The default Core-to-Core optimization pipeline contains, besides multiple invocations of the general simplifier, a few more specific passes: Common Subexpression Elimination (CSE), Demand analysis, Call Arity, Float-out and Float in, Worker-Wrapper transformation. Which of these do we want to run?

Most of them, such as the worker-wrapper transformation, make the code more efficient, but also more complex, and we do not expect them to be useful in these kind of proofs. Therefore, the small pipeline set up by the simplifier consists of multiple runs of the simplifier, interspersed with CSE.

One run of the simplifier can be configured by its *phase*, a small number that is counted down towards zero. Phases are used by complex arrangements of rewrite rules such as list fusion ([5]) to order rules. We run the simplifier in phases 2 and 1. We do not run a phase 0 because in this phase, recursive functions like map get inlined, which does not enable any further interesting transformations, but makes the output of a failed proof harder to understand.

The main purpose of the CSE run is to simplify useless case analyses expressions like **case** e **of** Succs x xs → Succs x xs to e.

*2.3.3 Aggressive inlining.* Inlining, i.e. replacing a function call with the body of the called function, is crucial for the effectiveness of this approach. GHC is generally very good at inlining, supporting, for example, cross-module inlining. But it refrains from inlining of larger functions when it does not see an obvious performance benefit, to avoid the program code size to explode.

In our case, however, there are more reasons for inlining than just performance, and code size is much less an issue. Therefore the plugin instructs the simplifier to be very aggressive when deciding whether to inline a function, and do not heed code size any more. Other ways to control inlining, such as the function arity, NOINLINE pragmas and phase control, are still adhered to.

## 3  WHAT CAN WE PROVE?

So far so simple. How far does this approach get us?

### 3.1  Type class instances

The type class instances in Figure 1 come with 9 laws. Of these, 7 laws go through immediately.

One failing law is

monad_law1 :: a → (a → Succs b) → Proof
monad_law1 a k = (return a ≫ k) === k a

which fails with

```
GHC.Proof: Proving monad_law1 ...
Proof failed
    Simplified LHS:
        Succs.Succs
         @ b
```

```
       (case k a of { Succs y ys -> y })
       (case k a of { Succs y ys -> ys })
Simplified RHS: k a
```

```haskell
module Bag where
data Bag a = Empty
          | Singleton a
          | Append (Bag a) (Bag a)
mapBag :: (a → b) → Bag a → Bag b
mapBag f Empty            = Empty
mapBag f (Singleton x)    = Singleton (f x)
mapBag f (Append b1 b2) = Append (mapBag f b1)
                                  (mapBag f b2)
append Empty b = b
append b Empty = b
append b1 b2 = Append b1 b2
```

**Figure 5: The Bags data structure**

Upon closer inspection we can see that the two sides differ only in strictness and evaluation order: While the LHS is always a value and calls k a only if the fields of Succs are evaluated, the RHS immediately calls k a.

This can mean one of two things to the user: Either he learns that he indeed did not write a law-abiding monad instance, and may have to revise the definition. Or he decides that fast-and-loose reasoning [3] is fine and he is not worried about differences in the strictness. He can even encode this decision in the proof obligation and change it to

monad_law1 :: a → (a → Succs b) → Proof
monad_law1 a k = (k a `seq` return a ≫ k) === k a

where the invocation of seq forces the left hand side to be strict in k a. The modified equality is proven by our plugin.

A similar problem arises with the other failing proof obligation, namely the third monad law:

monad_law3 :: Succs a → (a → Succs b)
              → (b → Succs c) → Proof
monad_law3 m k h = m ≫ (λx. k x ≫ h) ===
                    (m ≫ k) ≫ h

This law only holds if h is strict, so in order for this to pass, the user would instead write

monad_law3 m k h = m ≫ (λx. k x ≫ (h $!)) ===
                    (m ≫ k) ≫ (h $!)

to force h to be strict in its argument.

### 3.2  Custom data type

The code in Figure 1 in the previous section makes heavy use of the built-in data type for lists. Were the proof in the previous section maybe only successful because the compiler handles handle list functions specially?

To test that, changed replaced the use of lists in Figure 1 with *bags*, which ordered collections represented by trees.[1] The Succs

---
[1] In fact, due to the $O(1)$ concatenation, bags are more a suitable data structure here anyways.

```
{-# NOINLINE [0] mapBag # -}
{-# RULES "map/id" mapBag (λx. x) = id # -}
{-# RULES "map/." forall f g b. mapBag f (mapBag g b) = mapBag (f ∘ g) b # -}
{-# RULES "map/Empty" forall f. mapBag f Empty = Empty # -}
{-# RULES "map/append" forall f b1 b2. mapBag f (b1 `append` b2) = mapBag f b1 `append` mapBag f b2 # -}

{-# NOINLINE [0] append # -}
{-# RULES "append/Empty1" forall b. Empty `append` b = b # -}
{-# RULES "append/Empty2" forall b. b `append` Empty = b # -}
{-# RULES "append/assoc" forall b1 b2 b3. (b1 `append` b2) `append` b3 = b1 `append` (b2 `append` b3) # -}
```

Figure 6: Domain-specific rewrite rules

code needs to construct empty bags, map functions over bags and append bags; these operations are implemented as Figure 5.

After switching to this data type, all proof obligations fail. So clearly, the compiler knows something about list operations that he does not know here. But that is not a problem: We can simply tell the compiler what he needs to know, in the form of rewrite rules. These rules, listed in Figure 6, can be shipped with the Bag code and will be applied whenever the compiler sees code that matches the left-hand side of such a rule.

The NOINLINE pragmas are required to prevent the compiler from inlining these functions before the rules had a change to be applied.

Rule "append/assoc" is especially interesting: The append operator is not really associative, as it is easy to find bags b1, b2 and b3 so that the expression on the left-hand side of the rule can be distinguished from the right-hand side of the rule. But both sides represent the same collection, and all functions over bags are expected to respect this. In this sense, the equation holds *morally*, and we, as the library designer, may want to allow the compiler to perform this transformation. Since this is domain-specific knowledge, we cannot expect even the smartest compiler to do this on its own.

With these rules in place all 9 type class laws are proven again.

### 3.3 HLint rules

The HLint tool analyses Haskell code and suggests improvements to it. It has a database of patterns and replacement, which has been the target of formal verification before [1]. At the time of writing, we looked at 92 proof obligations, of which GHC is able to prove 34.

It is notable that even expressions involving IO are within reach. The plugin can prove, for example, the following identity:

proof1 x = putStrLn (show x) === print x

## 4 DISCUSSION

Let us return to the questions from the introduction.

(1) We have seen that it is possible to specify these proof obligations without extending the surface syntax of Haskell. Nevertheless, it might be desirable to do so in the long run, as the current approach has a few shortcomings. If the plugin does not "find" the (===), a proof obligation might simply be unnoticed. This might happen if the user abstracts over this operator.

(2) We check the output at the Core level, where all interesting transformations have been applied, but the expressions still resemble the user's code. Currently we compare the expressions for alpha-equality, but this can be too strict. One would reasonably expect the two expressions

λx y. **case** x **of** (x2, _) → **case** y **of** (y2, _) → (x2, y2)

and

λx y. **case** y **of** (y2, _) → **case** x **of** (x2, _) → (x2, y2)

to be considered equal. A more liberal equivalence relation on Core terms could allow that – and would be useful in other places in the compiler, e.g. in CSE.

(3) The previous section shows that indeed non-trivial equalities are proven by this approach. If the compiler is not sufficiently smart yet, the programmer is able to educate it about further equalities.

GHC's rewrite rule system is already quite useful, but still rather simple compared to the systems in theorem provers, e.g. Isabelle's simp method. If more sophisticated term rewriting features are added to GHC to meet a demand created by this application, then Haskell library authors would also benefit from the additional power.

(4) Of course, this approach will never disprove an equality. If the compiler tells the user that it cannot prove an equality, it very likely means that it is simply out-of-reach for this approach. The user may investigate the given terms and determine if they ought to be equal and can add some some strictness annotations or additional rewrite rules.

(5) The trustworthiness of these "proofs" raise some interesting questions. On the one hand we can see the compiler as a huge pile of unverified and bug-infested code, and we have no formal guarantees that the compiler respects some implicit or explicit semantics of the code. So the compiler saying that an equality holds does not mean anything.

On the other hand, we can see the compiler as the practically relevant way of giving semantics to our code in the first place. And with regard to this implementation-defined semantics, the compiler is correct by definition!

(6) Asking the compiler questions about how he compiles certain expressions has useful applications beyond proving theorems to be equal.

When developing a well-performing Haskell libraries, e.g. the list functions in the base library or the vector library, the library authors pay close attention to how the compiler compiles certain combinations of the library's functions. This is especially true when the Libra ships rewrite rules, as reading the Core output is the only way to make sure that the rules have the desired effect.

But when a new version of the compiler is released, it is not unlikely that it compiles the code in question slightly differently, and suddenly the optimizations do not happen in the intended way any more. But unless the library author manually checks the output, or has performance regression tests sensitive enough to notice the difference, this might go unnoticed.

But using the methods described here, the library author can explicitly list both the high-level code that it wants its user to be able to write and the low-level code that this should be compiled to (or rather, the Haskell source equivalent to the expected Core code), and ensure that this happens as part of the library's regular test suite.

In the context of test suites there is another interesting application of this idea. Imagine that the author of the Succs library would have used QuickCheck to test the type class laws instead. Whenever the the test suite is run the test suite would generate a hundred random inputs, pass them to the properties and compare the outcome left and right of the === operator. But, as we have seen now, after compilation, the exact same code is on both sides of the comparison!

In that case it would speed up the test suite to skip generating arguments for the property in the first place. Moreover, the test suite could list the test as "statically known to succeed", which may convey useful information to the programmer – maybe he expected to test a tricky property and a vacuously satisfied property indicates a bug in the test suite?

## 5  CONCLUSION

We find that GHC's is able to prove non-trivial program equalities just by virtue of its optimizer, and with – some convincing – is able to allow users to make use of this ability. We do not expect this method to succeed only for a limited class of equations, but it is unclear yet just how small or large this class is. Increasing the reach of the method may require more configurability (i.e. selectively disable or enable rules). Furthermore, every failed proof may point to a way in which the compiler's optimizer itself can be extended, which may in turn improve the compiler itself. Finally, even if the proving aspect of this work remains a cute hack, the outlined applications in testing may have actual impact.